\documentclass[runningheads]{llncs}
\usepackage{graphicx}
\usepackage[section]{placeins}
\usepackage[title]{appendix}
\usepackage{multirow}
\usepackage{subfig}
\begin{document}
\title{A Hierarchy-based Analysis Approach for Blended Learning: A Case Study with Chinese Students}
\titlerunning{A Hierarchy-based Analysis Approach}

\author{Yu Ye\inst{1}\and Gongjin Zhang\inst{2}\and Hongbiao Si\inst{3}\and Liang Xu\inst{3}\thanks{Corresponding author: Liang Xu, xuliang@hnchasing.com}\and
Shenghua Hu\inst{1}\and Yong Li\inst{2}\and Xulong Zhang\inst{4}\and Kaiyu Hu\inst{5}\and Fangzhou Ye\inst{6}}
\authorrunning{Y. Ye et al.}
%
\institute{Chasing Jixiang Life Insurance Co., Ltd., China\and Hunan Chasing Digital Technology Co., Ltd., China\and Hunan Chasing Financial Holdings Co., Ltd., China\and Ping An Technology (Shenzhen) Co., Ltd., China\and Stony Brook University, USA\and Chinasoft Co., Ltd., China }

\maketitle              
\begin{abstract}
Blended learning is generally defined as the combination of traditional face-to-face learning and online learning. This learning mode has been widely used in advanced education across the globe due to the COVID-19 pandemic's social distance restriction as well as the development of technology. Online learning plays an important role in blended learning, and as it requires more student autonomy, the quality of blended learning in advanced education has been a persistent concern. Existing literature offers several elements and frameworks regarding evaluating the quality of blended learning. However, most of them either have different favours for evaluation perspectives or simply offer general guidance for evaluation, reducing the completeness, objectivity and practicalness of related works. In order to carry out a more intuitive and comprehensive evaluation framework, this paper proposes a hierarchy-based analysis approach. Applying gradient boosting model and feature importance evaluation method, this approach mainly analyses student engagement and its three identified dimensions (behavioral engagement, emotional engagement, cognitive engagement) to eliminate some existing stubborn problems when it comes to blended learning evaluation. The results show that cognitive engagement and emotional engagement play a more important role in blended learning evaluation, implying that these two should be considered to improve for better learning as well as teaching quality. 

\end{abstract}
\section{Introduction}
Blended learning, commonly defined as “the integration of traditional face-to-face learning and online teaching” \cite{garrison2004blended,bliuc2007research,boelens2015blended}, has increasingly gained popularity and been widely implemented in higher education across the world. This process was greatly accelerated by the COVID-19 pandemic and the following global social distance restriction \cite{mahaye2020impact}. During this difficult period, remote learning has become common in students routine \cite{sun2022Pre-Avatar}. Besides, teleconferencing tools like Zoom help the delivery of online seminars and lectures, making remote education practical and popular. However, virtual learning, which mainly consists of online instruction and classes, is not diminished with the over of the pandemic and social distance restriction. In fact, remote learning is still an important part of the courses and programmes in higher education. Besides, profiting by the advancement of technology, this learning delivery mode is anticipated to continually be the mainstream in future higher education \cite{castro2019blended}. Therefore, the high-quality of online or blended education needs to be guaranteed.

The successful implementation of blended learning requires effective combination of virtual as well as face-to-face instruction \cite{garrison2004blended} rather than solely adding virtual learning elements, and this is not easily achieved. The reason is that different from face-to-face learning, remote learning often suffers from the lack of presence, reducing student engagement and thus harming the quality of learning. To achieve success in blended learning, students' self-motivation, self-reliance, independent study skills \cite{wivell2015blended}, and online engagement \cite{reed2014staff,chen2015checkable} are considered equally vital. This indicates that blended learning has a higher demand on overall student engagement in order to ensure learning quality \cite{deakin2015developing}. To achieve this, ongoing evaluation is regarded as essential \cite{moskal2013blended}. On one hand, it is claimed that the introduction of blended learning should be rather cautious at first to permit suitable tutor training and student adaption \cite{boyle2003using}. This implies the importance and necessity of ongoing evaluation in this gradual adaptation process as evaluation encourages reflections and improvements, helping better implementation in the future. On the other hand, ongoing evaluation is believed to give a more thorough and multi-faceted insight of the quality of blended learning. This improvement is believed to be beneficial for the overall high-quality of teaching in turn \cite{pombo2012evaluation}. 

In literature, certain factors that should be taken into account while evaluating blended learning have been mentioned. Course outcomes \cite{lopez2011blended,kiviniemi2014effects}, learner satisfaction \cite{chen2016empirical}, and student engagement \cite{holley2010student,vaughan2014student} are typical key components, of which student engagement is regarded as a more comprehensive criterion than the others. Additionally, many scholars have found a general positive relationship between the quality of blended learning and student engagement \cite{delialiouglu2012student,saritepeci2015effect,sari2016empirical}, making this criterion an outstanding indicator in the evaluation of blended learning. In terms of evaluation frameworks, diverse of them have been established with varying aims, engaged roles, evaluation focus, and judgement criteria. However, no certain one has ever received widespread recognition as the most efficient. Meanwhile, typically investigated through questionnaires, interviews, or simple classroom observations, these frameworks are more qualitatively based, causing the problem of subjectivity. Moreover, while existing research have broadly analysed western students’ experience, scholars have paid little attention to Chinese higher education and provided bare insights, reducing the generalisability of existing conclusions. 

Inspired by these studies, we consider a quantitative evaluation of blended learning and propose a hierarchy-based analysis approach for evaluation, using Chinese students' experience as a case study. Our work focuses on the perceptions of students and uses student engagement as the main evaluation indicator. Dividing student engagement into three dimensions, a questionnaire with matrix questions is conducted to collect primary dataset. After that, the importance of each dimension of student engagement is extracted. Consequently, the quality of blended learning is evaluated through the Analytic Hierarchy Process (AHP). Our contributions are summarised as follows: 

1) To evaluate the quality of blended learning, we propose a hierarchy-based analysis approach, improving the objectivity and accuracy of evaluation. 

2) With little research providing an insight into Chinese higher education, we narrow the gap by using Chinese students' experience as a case study to deepen the understanding.

\section{Related Work}
\subsection{Elements Regarding Evaluating Blended Learning}
Different elements have been pointed out in literature to be taken into consideration in terms of the evaluation of blended learning. Generally, major elements include course outcomes, learner satisfaction, and student engagement. 

Course outcomes are typically measured through aspects such as grades, class attendance, and drop out rates. Existing research has found that effective implementation of blended learning is beneficial for the improvement of course outcomes \cite{lopez2011blended,kiviniemi2014effects}. This criterion alone, however, fails to convey a comprehensive picture of the quality of blended learning because it neglects student's feelings and attitudes. One example is that students' motivation and initiatives towards learning are not captured. Therefore, whether blended learning helps facilitate these is not evaluated, which is noted as an important aspect regarding evaluating instructional effectiveness \cite{liu2012measuring}. 

Learner satisfaction offers a different perspective from course outcomes on the evaluation of blended learning by focusing on students' perceptions. Commonly measured by conducting self-report questionnaires, this element not only consider assessment data, but also other aspects such as learning environment, course content and flexibility, and perceived ease use of technology \cite{asoodar2016framework}. Thus, it comprehensively reflects students' personal experience and overall satisfaction of blended learning. This element also is proved to be positively affected by effective blended learning \cite{chen2016empirical,prifti2022self}.

Student engagement enables a deeper comprehension of the effectiveness of blended learning as it captures the contribution that students make to learning process for desirable outcomes \cite{kuh2011piecing} and the degree to which they engage in high-quality educational activities \cite{krause2008students}. Three dimensions of student engagement are identified: cognitive engagement, behavioural engagement, and emotional engagement \cite{fredricks2004school}. Generally, behavioural engagement relates to students' actions, having some overlaps with course outcomes. This dimension is mainly measured by students' involvement in learning process, such as actively attending class, collaborating with group members, and interacting with faculty \cite{kuh2001national}. Emotional engagement emphasises students' affective attitudes towards learning, such as interest, enjoyment and satisfaction. Cognitive engagement is relevant to the psychological investment in learning, such as self-management, initiatives towards learning and critical comprehension of knowledge. Positively affected by and giving a more full picture of blended learning, student engagement is becoming a crucial indicator for evaluation \cite{dringus2013five,saritepeci2015effect,sahni2019does}. 

\subsection{Evaluation Frameworks}
Based on elements mentioned above, different frameworks have been developed to evaluate blended learning with various purposes, involved roles, and evaluation focus. However, not a particular one has been commonly regarded as the most effective. Some selected frameworks will be discussed in the following parts. 

\textbf{Web-Based Learning Environment Instrument (WEBLEI):}
This framework focuses on investigating students' perceptions of e-learning environments. Four scales are incorporated, including emancipatory activities (focusing on convenience of materials, learning efficiency, and autonomy), co-participatory activities (focusing on students' learning processes such as flexibility, reflection, quality, interaction, feedback and collaboration), qulia (focusing on learning attitudes like enjoyment, frustration and tedium), and information structure (focusing on the design and arrangement of learning content). The first three are developed from Tobin's qualitative evaluation of Connecting Cummunities Learning (CCL) \cite{tobin1998qualitative} , and the last one is separately proposed by Chang \cite{chang1999evaluating}.  

\textbf{Hexagonal E-Learning Assessment Model (HELAM):}
This is a multi-dimensional approach in terms of evaluating learning management systems, focusing on the perception of learner satisfaction. Evaluated through a questionnaire, it has six evaluation criteria: system quality, information (content) quality, instructor attitudes, supportive elements, service quality and leaner perspective \cite{ozkan2009multi}. All of these dimensions are demonstrated to be significant. However, neglecting perspectives of other stakeholders, this model is questioned to some extent for only focusing on students. 

\textbf{E-Learning framework:}
 This is also a multi-dimensional framework containing eight systemically interconnected dimensions. They are technological (looking at infrastructure planning), pedagogical (looking at the arrangement and design of learning materials as well as learning strategies), interface design (looking at content design, navigation, and usability testing), evaluation (looking at learner assessment and teacher instruction), management (looking at maintaining learning environment and information transfer), resource support (looking at required remote support and resources), ethical (looking at social and ethical issues), and institutional (looking at administrative affairs and students services) \cite{deegan2015potential,gomes2015teaching}. However, instead of proposing any evaluation instrument, it only offers guidance for evaluating the environments of blended learning. 

\textbf{Rubric-based frameworks:}
This kind of frameworks have been created by several researches, commonly relying on judgement and having wide-ranging scopes. Evaluation factors mainly include instructional design, technology utilisation as well as students' experiences. Besides, these frameworks offer a quick and efficient method in terms of course evaluation for programme designers. Rubric-based frameworks are argued to be practically employed \cite{smythe2011blended}. However, depending heavily on judgements, these frameworks are inherently subjective. Additionally, not offering guidance for making judgements, evaluation provided by rubrics is judged to be broad and lacking depth. 

\section{Method}
We propose a hierarchy-based analysis approach to evaluate the quality of blended learning mainly based on the importance of all features to three dimensions of student engagement. To target Chinese students’ blended learning experience in higher education, an online survey was firstly created, measuring each feature with matrix questions with a seven-pointed Likert scale. Also, this survey was adapted from existing surveys in order to increase validity. Besides, previous studies find that gender \cite{kinzie2007relationship,lietaert2015gender} and age \cite{gibson2010student} both have an impact on student engagement. Therefore, they are also set as features to avoid potential bias. In terms of the measurement of blended learning, existing studies point out that effective mixture of face-to-face and virtual learning rather than simple adding virtual course materials constitutes a sufficient blend \cite{garrison2004blended}. This paper consider 30\%-80\% as an appropriate proportion of online learning contributing to blended learning \cite{allen2007extent}. Table \ref{table_measurement} summarises the main features, targets and their measures.
\begin{table}[]
\caption{Main features, targets and their measures}\label{table_measurement}
\resizebox{\linewidth}{!}{
\begin{tabular}{clccccccccccc}
\hline
\multicolumn{2}{c}{Category} &
  Measure & Matrix focuses\\ \hline
\multicolumn{2}{c}{\multirow{3}{*}{Behavioural Engagement (BL)}} &
  Active involvement (B-Act) & Attendance, Seats, Attention, Notes, Duration\\
\multicolumn{2}{c}{} &
 Faculty interaction (B-Int) & Questions, Eye-contact, Reflection\\  
 \multicolumn{2}{c}{} &
 Group collaboration (B-Gro) & Discussion, Communication, Presentation\\
\multicolumn{2}{c}{\multirow{2}{*}{Cognitive Engagement (CE) }} &
  Self-management (C-Mgt) & Pre-reading, Revision, Time schedule\\
\multicolumn{2}{c}{} &
  Comprehension (C-Com) & Grades, Assignments, Critical thinking, Strategies\\ 
\multicolumn{2}{c}{\multirow{2}{*}{Emotional Engagement (EE) }} &
  Interest (E-Int) & Motivation, Related reading,  Inspiration
  \\
\multicolumn{2}{c}{} &
  Satisfaction (E-Sat) & Support, Confidence, Accomplishments, Enjoyment
 \\ 
 \multicolumn{2}{c}{\multirow{1}{*}{Blended Learning (BL) }} &
  Proportion of online learning &\
  \ \ \textbackslash{} \
  \\ \hline
\end{tabular}
}
\end{table}

\begin{figure}
\includegraphics[width=\textwidth]{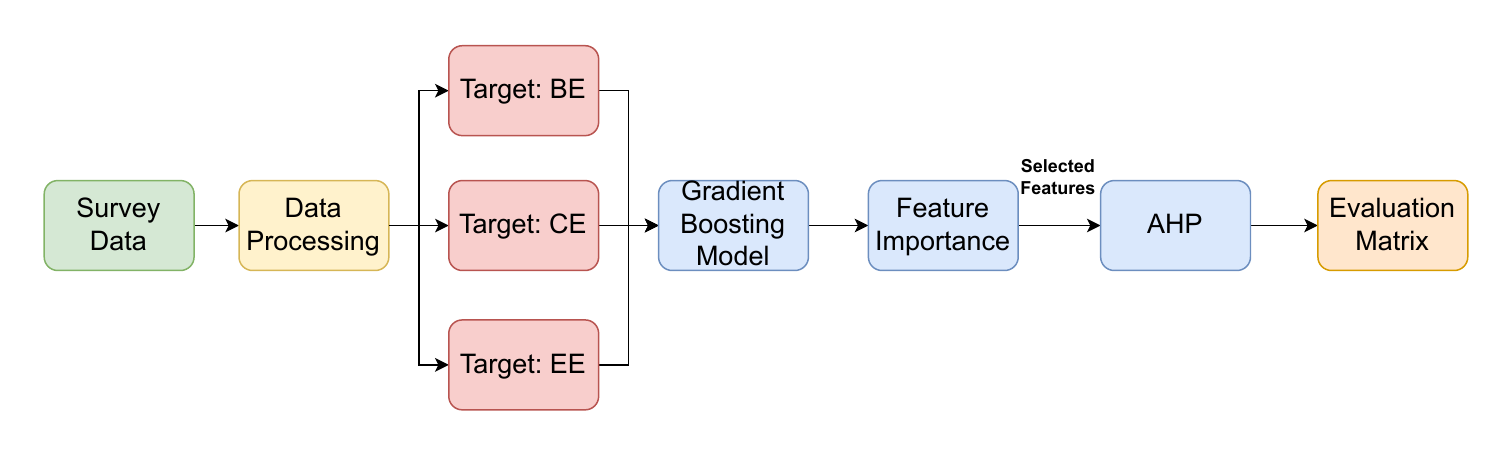}
\caption{The framework of AHP analysis approach. Gradient boosting regression model and feature importance analysis method are applied to extract important features. The importance value of each feature is then fed into the AHP method to calculate the evaluation matrix.} \label{fig_method}
\end{figure}

To collect primary dataset, our survey was spread through the online advanced education communities provided by Weibo, one popular Chinese social media. After that, gradient boosting regression model was applied to fit the survey data. Besides, gini importance and permutation importance were used to measure the importance of each feature to the regression target. Based on features selected, the analytic hierarchy process (AHP) method was then applied to build a evaluation matrix to measure student engagement. Fig. \ref{fig_method} presents the whole framework. 

Cognitive Engagement (CE), Behavioural Engagement (BE), and Emotional Engagement (EE) are separately set as the regression target $Y$. Each of the training/testing set was then fed into the Gradient Boosting Regression model.

\subsection{Gradient Boosting Regression}
Gradient Boosting Regression, also known as Gradient Boosted Decision Trees, is a model that can be applied to both classification and regression tasks. Compared to decision tree model or other simple linear models, it is capable of handling continuous features and discrete features. Besides, based on decision tree model, this model is relatively easy to fit and fine-tuning~\cite{friedman2002stochastic}. Our model takes a fixed-size decision tree as the weak learner and is built in a greedy manner:
\begin{equation}
\hat{y}_{i} = F_M(x_i) = \sum_{m=1}^M h_m(x_i)
\end{equation}
where $h_m$ is the set of decision tree model with size of $M$, also known as weak learners in the case of boosting method. 

In each gradient step, a new decision tree $h_m$ is added into the whole model, updating the $F_m(x)$ in the following greedy way:
\begin{equation}
F_m(x) = F_{m-1}(x) + h_m(x)
\end{equation}
A decision tree is a model that applies non-parametric supervised learning method to achieve the regression goal. It contains a set of if-then-else decision rules that can learn from the data points to approximate the regression curve. The tree added in each step will learn from the training data and try to minimise the losses function, which is the mean squared error function in this case: 
\begin{equation}
MSE(y, \hat{y}) = \frac{1}{n}\sum_{n-1}^{i=0}(y_i - \hat{y_i})^2
\end{equation}
According to Friedman \cite{friedman2002stochastic}, the decision tree $h_m$ predicts the negative gradients of the training data updated at each training step. The Gradient Boosting Regression can be regarded as a process of doing gradient descent in a functional space.

\subsection{Gini Importance and Permutation Importance}
Gini importance and permutation importance are used in feature importance area to measure the relevance between features and targets. 

Gini importance, also known as Mean Decrease Impurity (MDI), is a impurity-based method and represents the average and variability of impurity reduction accumulation within each individual tree~\cite{li2019debiased}. It is calculated as the following:
\begin{equation}
MDI(k, T) = \sum{\frac{N_n(t)}{n}\Delta{x(t)}} 
\end{equation}
where $X$ is the feature and $T$ is the weak learner.

The result of gini importance may be biased when the feature has a large amount of unique values. Therefore, the permutation importance is used as an alternative to overcome this. It calculates feature importance by evaluating the change in the model's performance when randomly permuting the values of a single feature~\cite{zhang2012ensemble}:
\begin{equation}
i_j = s - \frac{1}{K}\sum_{k=1}^{K}s_{k,j} 
\end{equation}
Where $i_j$ is the importance of feature $f_j$, $s$ is the reference score of the model on the dataset, and $K$ is the total repetition used to calculate the importance.

Both gini importance and permutation importance methods are used to evaluate the feature for better confidence level of feature importance.

\subsection{Analytic Hierarchy Process}
Analytic Hierarchy Process (AHP) is an effective method involving both qualitative and quantitative analysis \cite{saaty1987analytic}. It uses a hierarchy structure to divide the decision process into three levels - Alternatives, Criteria and Goal~\cite{lipovetsky1996synthetic}. The feature importance extracted from gini importance and permutation importance method is applied to initialise the pairwise comparison matrix. 

Table \ref{table_ahp} shows the matrix used in AHP to assign the intensity of importance to each criterion. The pairwise comparison is then established, and AHP will check the consistency. If the check is pass, the AHP method will output a weighted score for each criterion. 
\begin{table}
\begin{center}
\caption{AHP Comparison Index}\label{table_ahp}
\begin{tabular}{cc}
\hline
{\bfseries Intensity of importance} & {\bfseries Definition}\\
\hline
1 & Equal\\
2 & Weak\\
3 & Moderate\\
4 & Moderate plus\\
5 & Strong\\
6 & Strong plus\\
7 & Demonstrate\\
8 & Demonstrate plus\\
9 & Extremely preferred\\
\hline
\end{tabular}
\end{center}
\end{table}

As the pairwise comparison could be inaccurate due to user's subjective bias, the importance conducted from gini importance and permutation importance methods is therefore applied to reduce this. A mapping is created to map the importance learnt by the model to the pairwise comparison of the AHP method. The details will be discussed in the following experiments section.

\section{Experiments and Results}
\subsection{Dataset}
1132 samples were submitted to our online survey. Gender distribution shows that 69.3\% of the respondents are identified as females, while 30.7\% are identified as males. This gender imbalance implies that the interpretation of results may primarily reflect the experiences and perspectives of female participants, limiting the generalisability of the findings. In terms of the age distribution of the sample, with over 90\% of the sample's participants being over the age of 18, it ensures a representative sample that aligns with the target population under investigation. Furthermore, it is worth noting that most respondents (60\%) are between the ages of 18 and 21, indicating that the findings are more representative of undergraduate experience. 

Table \ref{table_dataset_details} provides a summarised overview of the descriptive statistics derived from the dataset. The mean values reveal that less than half of the student participants reported having experienced blended learning, which appears contrary to the prevailing trend of increased integration of online learning with conventional educational practices in light of the Covid-19 pandemic. A plausible explanation for this observation could be attributed to the varying extent to which online learning is embraced by individual students. With some demonstrating an excessive reliance on online platforms while others exhibiting a minimal incorporation of such methods into their overall learning routine, these fail to meet the specific criterion outlined for blended learning in this paper. Additionally, the average scores for each aspect of student engagement slightly surpasses 4, indicating a generally positive inclination towards active participation in educational activities within the blended learning environment. The relatively low standard deviations observed further suggest a convergence of responses around the mean values, implying a degree of consensus among the participants.

\begin{table}
\begin{center}
\caption{Descriptive statistics of dataset}
\label{table_dataset_details}
\resizebox{\linewidth}{!}{
\begin{tabular}{cccccccccccccc}
\hline
{\bfseries } & {\bfseries BL} & {\bfseries B-Act} & {\bfseries B-Int} & {\bfseries B-Gro} & {\bfseries C-Mgt} & {\bfseries C-Com} & {\bfseries E-Int} & {\bfseries E-Sat} & {\bfseries BE} & {\bfseries CE} & {\bfseries EE}\\
\hline
Mean & 0.4488 & 4.6693& 4.6614 & 4.5748 & 4.6457 & 4.4803 & 4.8661 & 4.669 & 4.6352 & 4.5630 & 4.7677\\
Std. D & 0.4993 & 1.5688 & 1.5287 & 1.6548 & 1.7207 & 1.6755 & 1.7922 & 1.7820 & 1.4496 & 1.6439 & 1.7410\\
Stewness & 0.2083 & -0.4756 & -0.4131 & -0.4166 & -0.3273 & -0.2687 & -0.6362 & -0.4709 & -0.4950 &-0.2793&-0.5443\\
Kurtosis & -1.9882 & -0.2727 & -0.2911 & -0.5827 & -0.7437 & -0.6816 &-0.4587&-0.6255&-0.1736&-0.6633&-0.5578\\
\hline
\end{tabular}    
}
\end{center}
\end{table}

\subsection{Experimental Setup}
The collected survey dataset is divided into a 80/20 split for training and testing the Gradient Boosting Regression model. Cognitive Engagement (CE), Behavioural Engagement (BE), and Emotional Engagement (EE) are set as the regression target $Y$ separately as shown in Table \ref{table_dataset_setup}. 
\begin{table}
\begin{center}
\caption{Data samples setup}\label{table_dataset_setup}
\begin{tabular}{cc}
\hline
{\bfseries Target $Y$} & {\bfseries Features $X$}\\
\hline
CE & Gender, Age, BL, B-Act, B-Int, B-Gro, E-Int, E-Sat, BE, EE\\
BE & Gender, Age, BL, C-Mgt, C-Com, E-Int, E-Sat, CE, EE\\
EE & Gender, Age, BL, B-Act, B-Int, B-Gro, C-Mgt, C-Com, BE, CE\\
\hline
\end{tabular}
\end{center}
\end{table}

The mean squared error is used as the loss function to train the gradient boosting model for 500 boosting stages with learning rate being 0.01. The max depth of each decision tree of the weak learner is 4. For evaluation, the training and testing deviance is applied to measure the learning process.

\subsection{Results and Analysis}
We first inspect the training and testing deviance of each dataset. At this stage, all parameters of three models and training process are set as the same except the dataset itself.

\begin{figure}[]
  \centering
  \subfloat[Target: BE] 
  {
      \label{fig:subfig1}\includegraphics[width=0.3\textwidth]{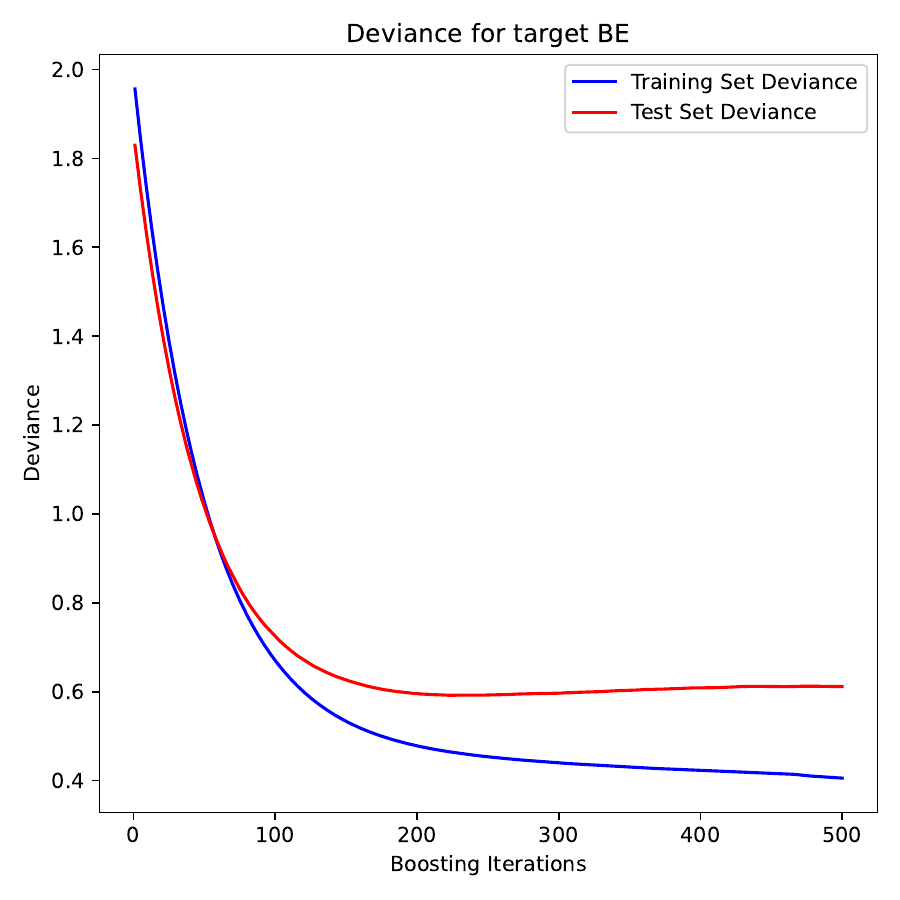}
  }
  \subfloat[Target: CE]
  {
      \label{fig:subfig2}\includegraphics[width=0.3\textwidth]{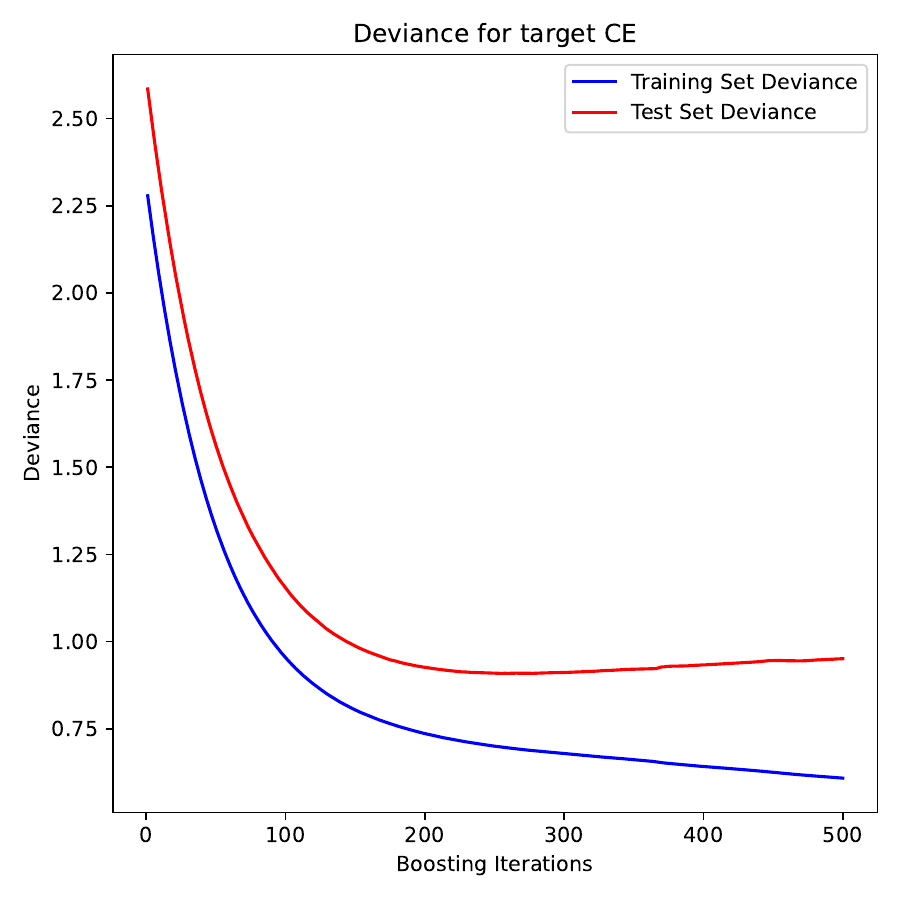}
  }
  \subfloat[Target: EE]
  {
      \label{fig:subfig3}\includegraphics[width=0.3\textwidth]{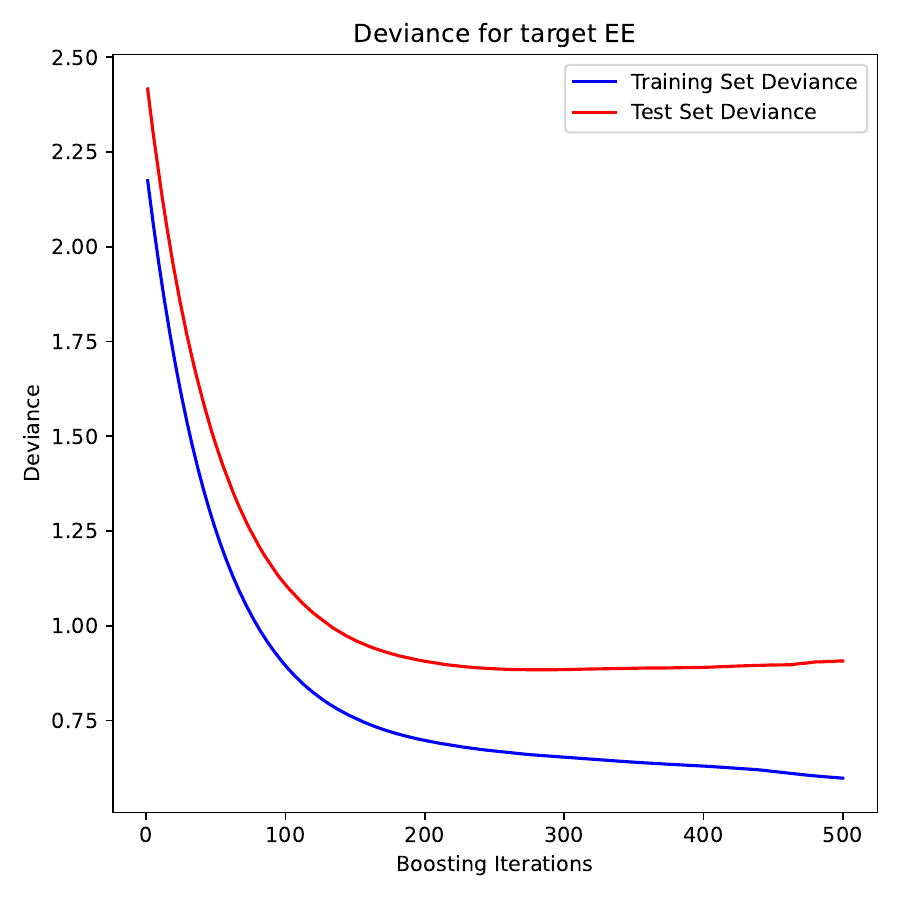}
  }
  \caption{The training and testing deviance on three datasets with different target $Y$. Fig. 2(a) is the model trained on the regression of target BE. Fig. 2(b) is the model trained on the regression of target CE. Fig. 2(c) is the model trained on the regression of target EE. All three models achieve the saturation point around iteration 200.}
  \label{training_result} 
\end{figure}

Fig. \ref{training_result} indicates that all models achieve the saturation point around 200 boosting iterations, meaning that every single model is capable for learning certain level of the representation from the training dataset. More iterations may lead to overfitting issue. In general, the results prove that feature importance conducted from this model is accurate and can be applied to the AHP method later on. 

The next step is to measure which features are more relevant to target $Y$. The feature importance results for all three models are similar, therefore we only illustrate regression model with target BE in detail.

\begin{figure}
\includegraphics[width=\textwidth]{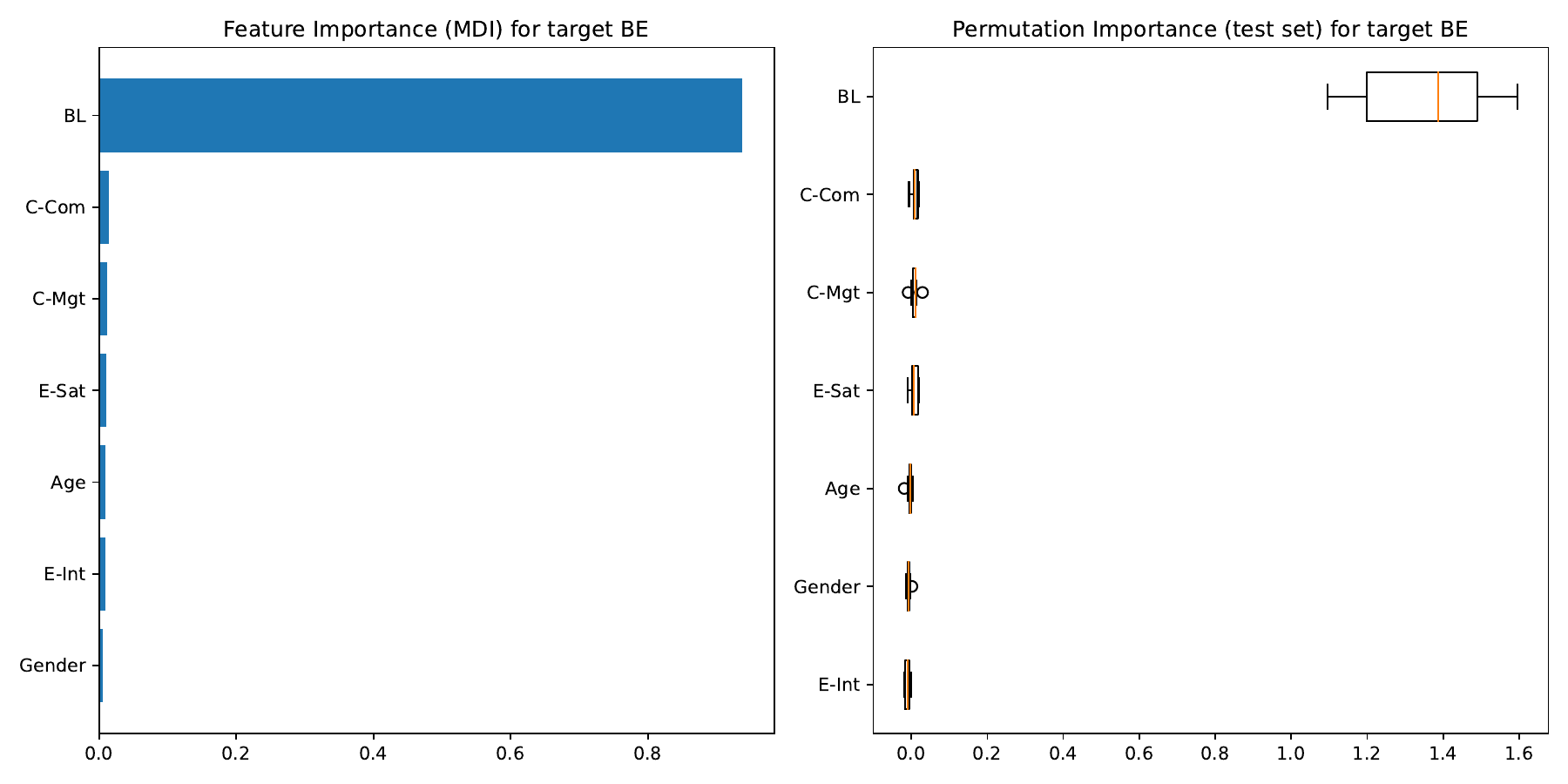}
\caption{Gini importance and Permutation importance of BE model. It shows that the BL feature has the most significant contributions to the target BE score and the rest features share similar level of importance. } \label{fig_feature_importance_BE}
\end{figure}

Fig. \ref{fig_feature_importance_BE} indicates that both gini importance and permutation importance show that the BL feature is the most relevant one. Similar conclusions can be drawn from the results of feature importance of other two models.

Finally, we map the gini importance and permutation importance to the pairwise comparison scale in the AHP method. It is clearly that the $BL$ feature is the demonstrated importance and $Age$ is the least significant feature. Therefore, the scale for $BL$ to other features is set as 7, the scale for $Gender$ to $BL$ is set as $1/9$, and the scale for other features is 3. The pairwise comparison matrix for the $BE$ model can be created as Table \ref{table_AHP_Pairwise}.
\begin{table}
\begin{center}
\caption{AHP Pairwise Comparison Matrix For BE.}\label{table_AHP_Pairwise}
\begin{tabular}{cccccccc}
\hline
{\bfseries Feature} & {\bfseries BL}& {\bfseries C-Com}& {\bfseries C-Mgt}& {\bfseries E-Sat}& {\bfseries Age}& {\bfseries E-Int}& {\bfseries Gender}\\
\hline
BL & 1 & 7 & 7 & 7 & 7 & 7 & 9\\
C-Com & 1/7 & 1 & 1 & 1 & 1 & 1 & 3\\
C-Mgt & 1/7 & 1 & 1 & 1 & 1 & 1 & 3\\
E-Sat & 1/7 & 1 & 1 & 1 & 1 & 1 & 3\\
Age & 1/7 & 1 & 1 & 1 & 1 & 1 & 3\\
E-Int & 1/7 & 1 & 1 & 1 & 1 & 1 & 3\\
Gender & 1/9 & 1/3 & 1/3 & 1/3 & 1/3 & 1/3 & 1\\
\hline
\end{tabular}
\end{center}
\end{table}

The square root method of AHP is applied to calculate the evaluation matrix and normalise weighted value for each feature. The final consistency index is $0.013$, meaning that the final matrix is consistent and the evaluation matrix conducted by AHP approach is valid. Similar approach can be applied to the model with the other two targets, and the final evaluation matrix is shown in Table \ref{table_evaluation_result}. It indicates that blended learning significantly affects student's BE, EE and CE in a positive way. Age is the least significant feature for all the three models, and other features share similar level of importance.
\begin{table}[]
\caption{AHP Evaluation Matrix for target BE, CE and EE.}\label{table_evaluation_result}
\resizebox{\linewidth}{!}{
\begin{tabular}{clccccccccccc}
\hline
\multicolumn{2}{c}{\textbf{Target}} &
  \textbf{Weight} &
  \textbf{BL} &
  \textbf{B-Act} &
  \textbf{B-Int} &
  \textbf{B-Gro} &
  \textbf{C-Mgt} &
  \textbf{C-Com} &
  \textbf{E-Int} &
  \textbf{E-Sat} &
  \textbf{Gender} &
  \textbf{Age} \\ \hline
\multicolumn{2}{c}{\multirow{2}{*}{BE}} &
  Weight Score &
  \textbf{5.495} &
  \textbackslash{} &
  \textbackslash{} &
  \textbackslash{} &
  0.886 &
  0.886 &
  0.886 &
  0.886 &
  0.886 &
  0.333 \\
\multicolumn{2}{c}{} &
  Percentage(\%) &
  \textbf{53.566} &
  \textbackslash{} &
  \textbackslash{} &
  \textbackslash{} &
  8.637 &
  8.637 &
  8.637 &
  8.637 &
  8.637 &
  3.249 \\ \hline
\multicolumn{2}{c}{\multirow{2}{*}{CE}} &
  Weight Score &
  \textbf{5.759} &
  0.981 &
  0.981 &
  0.981 &
  \textbackslash{} &
  \textbackslash{} &
  0.981 &
  0.981 &
  0.333 &
  0.574 \\
\multicolumn{2}{c}{} &
  Percentage(\%) &
  \textbf{49.733} &
  8.478 &
  8.478 &
  8.478 &
  \textbackslash{} &
  \textbackslash{} &
  8.478 &
  8.478 &
  2.881 &
  4.957 \\ \hline
\multicolumn{2}{c}{\multirow{2}{*}{EE}} &
  Weight Score &
  \textbf{5.759} &
  0.981 &
  0.981 &
  0.981 &
  0.981 &
  0.981 &
  \textbackslash{} &
  \textbackslash{} &
  0.333 &
  0.574 \\
\multicolumn{2}{c}{} &
  Percentage(\%) &
  \textbf{49.733} &
  8.478 &
  8.478 &
  8.478 &
  8.478 &
  8.478 &
  \textbackslash{} &
  \textbackslash{} &
  2.881 &
  4.957 \\ \hline
\end{tabular}
}
\end{table}

Our work evaluates the generalizability of previous theories and close any potential research gaps by examining this relationship of belended learning and student engagement in the context of Chinese higher education. Although our work has proved some results from previous research, it is worth nothing that this paper define study programmes with 30\% to 80\% online learning as blended learning. To deepen the understanding of the quality of blended learning, it is suggested that the proportion of online learning could be studied at a more granular level.

\section{Conclusion}
In this work, we examine how different aspects of student engagement relate to the quality or effectiveness of blended learning. According to our results, it is clearly that blended learning significantly affects student engagement in a positive way, particularly cognitive engagement and emotional engagement. Using student engagement as a indication, it is safe to conclude that the quality or effectiveness of blended learning can be gauged indirectly. Besides, the findings suggest that the trend towards blended learning being the norm in future higher education will be beneficial to increase learning quality. Additionally, proposing the AHP Approach for blended learning evaluation, it shows that cognitive engagement and emotional engagement are more important for learning quality. However, in terms of properly allocating the percentage of remote learning, it is still unclear how to maximise the advantages of blended learning. Therefore, academics are urged to gain a thorough grasp of the effectiveness of diverse combinations of face-to-face learning and online learning to further this academic research and benefit remote education.

\bibliographystyle{splncs04}
\bibliography{ref}


\end{document}